\documentclass[twocolumn,aps,prl,showpacs,epsfig]{revtex4}
\usepackage{amsfonts}
\usepackage{amsmath}
\usepackage{amssymb}
\usepackage{graphicx}
\begin{document}

\title{Competing Orders and Superconductivity in the 
Doped Mott Insulator on the Shastry-Sutherland Lattice}

\author{Chung-Hou Chung}
\author{Yong Baek Kim}
\affiliation{Department of Physics, University of Toronto, Toronto, Ontario, 
Canada M5S 1A7}
\date{\today}

\begin{abstract}
Quantum antiferromagnets on geometrically frustrated lattices often
allow a number of unusual paramagnetic ground states. The fate of 
these Mott insulators upon doping is an important issue that may
shed some light on the high $T_c$ cuprate problem. We consider the 
doped Mott insulator on the Shastry-Sutherland lattice via the
$t$-$J$ model. The U(1) slave-boson mean field theory reveals
the strong competition between different broken symmetry states.
It is found that, in some ranges of doping, there exist 
superconducting phases with or without coexisting 
translational-symmetry-breaking orders such as the staggered flux 
or dimerization. Our results will be directly relevant to 
SrCu$_2$(BO$_3$)$_2$ when this material is doped in future.
\end{abstract}

\pacs{74.20.-z,74.20.Mn}

\maketitle

{\bf Introduction}:
The understanding of the high $T_c$ superconductivity in layered 
cuprates remains one of the central problems in correlated 
electron physics. Even though there is no consensus as to 
the solution of the puzzle, it is widely believed that the physics 
of the cuprates has much to do with the doped spin-$1/2$ 
Mott insulator. It was suggested by Anderson that the strong 
quantum fluctuations 
associated with the spin-$1/2$ and two dimensionality may lead to
a quantum liquid of spin singlets or the resonating valence 
bond (RVB) state\cite{anderson87}. 
Upon doping, it would become a superconductor 
when the holes are phase coherent. The undoped cuprate is,
however, antiferromagnetically ordered, not a RVB liquid. Nonetheless, 
the hopping processes of the doped holes strongly frustrate the N\'eel 
order and the resulting spin-disordered state may have substantial 
correlations of the RVB state. This would be more likely the case 
if the RVB state were close in energy to the antiferromagnetic ground 
state in the undoped system. 

It is, therefore, extremely interesting to identify 
the Mott insulators with no long-range spin order. 
It has been known for sometime that quantum antiferromagnets on 
the geometrically frustrated lattices are such examples\cite{sachdev92}. 
At the classical level, the frustration often leads to a large 
degeneracy of the classical ground state and the 
resulting fluctuations can suppress the classical long-range 
spin-order. Previous studies showed that various quantum
paramagnetic ground states can also occur at zero temperature
and they include various translational-symmetry-breaking 
phases as well as the RVB state\cite{sachdev92}. 
In this case, the doping
of holes or electrons may naturally lead to superconductivity.

In this context, the recent discovery of superconductivity
in layered cobaltates, Na$_x$CoO$_2$$\cdot y$H$_2$O, deserves 
particular attention\cite{takada03}. 
It has been suggested that this system 
can be regarded as an electron doped Mott insulator with 
spin-1/2 Co ions on the triangular lattice\cite{dhlee03}. 
Recent theoretical studies of a model based on the RVB 
picture found a superconducting 
phase with the $d+id$ pairing symmetry\cite{dhlee03,ogata03}. 
While the cobaltates is 
in a doped state, an undoped spin-1/2 Mott insulator, Cs$_2$CuCl$_4$, 
with Cu spins on the anisotropic triangular lattice, has been also
identified\cite{coldea01}. The ground state is magnetically ordered, 
but the system seems to be very close to the phase transition 
toward the RVB or spin-liquid state, as seen from the neutron 
scattering experiments\cite{coldea01} and a large-$N$ mean-field
theory\cite{chung03,zhou03}. 

These discoveries raise the hope that one may be able to
achieve ``high $T_c$'' superconductivity by doping frustrated
magnets. Moreover, the studies of these systems may provide
an insight about the role of the dynamic frustration 
due to the motion of doped holes in cuprates and its 
influence on the superconductivity, albeit it
is not the same as the geometric frustration.   

In this paper, we study possible superconducting states and other
competing orders in the doped Mott insulator on the Shastry-Sutherland 
lattice. The lattice structure is shown in Fig.1.
This work is strongly motivated by the discovery of the
Mott insulator SrCu$_2$(BO$_3$)$_2$ where the spin-1/2 Cu ions
lie in two dimensional layers decoupled from each other\cite{kageyama99}.
Remarkably, the topology of the nearest-neighbor antiferromagnetic 
exchange couplings between the Cu ions is identical to the Hamiltonian 
studied by Shastry and Sutherland many years ago\cite{shastry81}.
The model is the nearest-neighbor antiferromagnetic Heisenberg model 
on the Shastry-Sutherland lattice with different exchange couplings
on the square lattice links ($J$) and diagonal links ($J'$).
The quantum phase diagram of this model in a wide parameter range 
has been extensively studied recently\cite{koga00,chung01}.
Possible paramagnetic ground states include
the plaquette-ordered phase, decoupled-dimer state, and 
the topologically ordered RVB spin-liquid phase\cite{koga00,chung01}. 

In the case of SrCu$_2$(BO$_3$)$_2$, the values of the exchange 
couplings ($70-85 K$)\cite{ueda99} are much smaller than those 
in high-T$_c$ cuprates (about $1500 K$), we therefore expect a lower 
superconducting transition temperature for the putative superconducting
state at finite doping. The experiments on the undoped system show 
clear spin-gap behavior\cite{kageyama99} and the material at ambient
pressure may be in the decoupled-dimer state with the dimers 
on the diagonal links. This state is likely to be very 
close to the phase boundary of the spin-liquid state according to the 
large-$N$ mean field calculation\cite{chung01}. 
We expect, therefore, that there will 
be strong competitions between different orders upon doping.  

The appropriate model to describe this material at finite doping 
is the $t$-$J$ model on the Shastry-Sutherland lattice.
This model was previously studied by Shastry and Kumar\cite{shastry02} 
via a RVB mean-field approach where only the order parameters
in the particle-particle channel were considered. As has been done in
the cases of the square and triangular 
lattices\cite{dhlee03,palee92,kotliar88}, here we consider
both the particle-hole and particle-particle channels and their mutual 
influence. We used the U(1) slave-boson
mean field theory to get the phase diagram at zero and finite 
temperatures. It turns out that the consideration of the order parameters
in the particle-hole channel is very important and leads to a   
qualitatively different phase diagram. 

We start from the insulating decoupled dimer phase at the half-filling,
that is appropriate for SrCu$_2$(BO$_3$)$_2$. At low temperatures,
three superconducting phases appear successively as the doping is 
increased; a superconductor with dimerization, followed by a first
order transition to a $d$-wave superconductor coexisting with the
staggered flux order, and finally a uniform $d$-wave superconductor.
Details of the phase diagram and other competing phases will be
discussed below. We start with the description of the
slave-boson mean-field theory.

{\bf U(1) slave-boson mean-field theory}:
In the U(1) slave-boson formulation\cite{palee92,kotliar88}, 
the electron operator $c^{\dagger}_{i\alpha}$ 
can be written as  
$c^{\dagger}_{i\alpha} = f^{\dagger}_{i\alpha} b_{i}$
($\alpha = \uparrow, \downarrow$), where 
the fermion (boson) operator $f^{\dagger}_{i\alpha}$ ($b_i$)
carries the spin (charge) quantum number. In this case, the constraint 
$f^{\dagger}_{i\alpha}f_{i\alpha} + b^{\dagger}_{i}b_{i} = 1$
should be imposed to exclude doubly occupied sites in 
the Hilbert space. Then the $t$-$J$ model can be written as
\begin{eqnarray}
H &=& -\sum_{<i,j>} t_{ij} (f^{\dagger}_{i\alpha}f_{j\alpha} 
b^{\dagger}_{j} b_{i} + h.c.) \cr 
&+& \sum_{<i,j>}J_{ij}(\vec{S}_{i}\cdot \vec{S}_{j} 
- \frac{1}{4} n_{i} n_{j}) 
- \mu \sum_{i} f^{\dagger}_{i\alpha}f_{i\alpha} \cr
&+& i \sum_{i} \lambda_{i} 
(f^{\dagger}_{i\alpha}f_{i\alpha} + b^{\dagger}_{i}b_{i} - 1),
\end{eqnarray}
where $t_{ij}$ and $J_{ij}$ are the hopping parameters and 
antiferromagnetic exchange 
couplings such that $t_{ij} = t$, $J_{ij} = J$ on the
nearest-neighbor links and $t_{ij} = t'$, $J_{ij} = J'$ 
on the diagonal links (see Fig.1).
Here $\vec{S}_i = \frac{1}{2} f^{\dagger}_{i\alpha}
\vec{\sigma}_{\alpha\beta}f_{j\beta}$,
$n_{i} = f^{\dagger}_{i\alpha} f_{i\alpha}$,
and $\lambda_i$ is the Lagrange multiplier introduced to 
impose the constraint.

In order to obtain the mean-field theory, we introduce the
Hubbard-Stratonovitch fields ${\tilde \chi}_{ij}$ and $\Delta_{ij}$
to decouple the four-fermion interaction in the particle-hole
and particle-particle channels, respectively.
The resulting mean-field Hamiltonian can be 
written as\cite{palee92}
\begin{eqnarray}
H_{mf} &=& \sum_{<ij>} [ \frac{3J_{ij}}{8} (|{\tilde \chi}_{ij}|^2 
+ |\Delta_{ij}|^2 ) \cr
&-& {\tilde \chi}_{ij}^{*}(f^{\dagger}_{i\alpha}f_{j\alpha} + 
\frac{8t_{ij}}{3J_{ij}} b_{i}^{\dagger}b_{j}) 
- \Delta_{ij}^{*} \epsilon_{\alpha \beta}
f_{i\alpha}f_{j\beta} + h.c. ] \cr 
&-& \mu \sum_{i} f^{\dagger}_{i\alpha}f_{i\alpha}
- i \sum_{i} \lambda_{i}(f^{\dagger}_{i\alpha}f_{i\alpha} + 
b^{\dagger}_{i}b_{i}-1)
\end{eqnarray}
At the saddle point of the mean-field action, we get
\begin{eqnarray}
{\tilde \chi}_{ij} &=& \chi_{ij} +  
\frac{8t_{ij}}{3J_{ij}} \left < b_{i}^{\dagger}b_{j} \right > 
\ \ {\rm with} \ \ \chi_{ij} =  
\left < f^{\dagger}_{i\alpha}f_{j\alpha} \right > \cr 
\Delta_{ij} &=& \left < \epsilon_{\alpha \beta}
f_{i}^{\alpha}f_{j}^{\beta} \right >.
\end{eqnarray}
Notice that $\chi_{ij}$ can be used as an alternative
mean-field variables instead of ${\tilde \chi}_{ij}$.

\begin{figure}[h]
\includegraphics[height=5cm,width=5cm,angle=0]{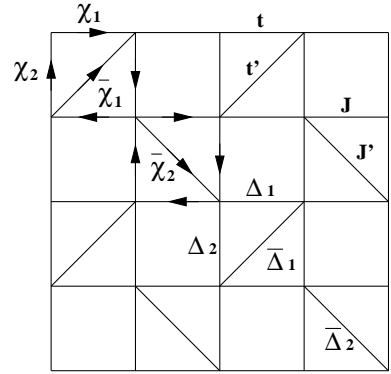}
\caption{The Shastry-Sutherland lattice. The parameters used in 
the $t$-$J$ model on this lattice are given by 
$t/t'= J'/J=1.05$, and $t'/J = 2.0$.
The eight different order parameters of the model 
are defined on the four-site 
unit cell: $\chi_1$ and $\Delta_1$ ($\chi_2$ and $\Delta_2$) fields 
reside on the horizontal (vertical) links; while $\bar{\chi}_1$,
$\bar{\chi}_2$, $\bar{\Delta}_1$, and $\bar{\Delta}_2$ live on the 
diagonals. The arrows represent the directions of the $\chi$ fields.
}
\label{lattice}
\end{figure}
 
At the half-filling, the original Hamiltonian has a hidden 
SU(2) symmetry\cite{affleck88}. Upon doping, this symmetry is broken down 
to U(1) and there exist a number of equivalent mean-field states that 
can be transformed into each other under the U(1) gauge transformation:
$f^{\dagger}_{i\sigma} \rightarrow f^{\dagger}_{i\sigma} e^{i\phi_i}$, 
$b^{\dagger}_{i} \rightarrow b^{\dagger}_{i} e^{i\phi_i}$, 
$\chi_{ij} \rightarrow \chi_{ij} e^{i(\phi_i - \phi_j)}$, and 
$\Delta_{ij} \rightarrow \Delta_{ij} e^{i (\phi_i + \phi_j)}$. 
Therefore, the mean-field states should be classified using the 
the gauge-invariant quantities. It will prove useful to utilize
$\prod_{ijkl} \chi_{ij}\chi_{jk}\chi_{kl}\chi_{li}$ on
each plaquette and $\prod_{ijk}\chi_{ij}\chi_{jk}\chi_{ki}$
on each triangular loop. The phases of these products 
can be interpreted as the ``flux''\cite{marston88} of the U(1) gauge field
going through a plaquette or a triangle. If the ``flux'' away from 
half-filling does not equal to $\pi$ or $0$, it will generate the 
staggered orbital current proportional to 
${\rm Im}(<b_{i}^{\dagger}b_{j}> \chi_{ij}^*)$ 
running through the plaquette or triangle.  
The superconductivity requires the condensation of the bosons
$<b_i>\not=0$ as well as the formation of the RVB singlet
$\Delta_{ij} \not= 0$.
Combined with these considerations, the mean-field phase diagram
can be determined by minimizing the mean-field free energy at 
a given temperature $T$ and a doping 
$\delta = \sum_{i}b^{\dagger}_{i}b_{i}
= 1 - \sum_{i}f^{\dagger}_{i\alpha}f_{i\alpha}$.
In our study, we use $t/t'= J'/J = 1.05$, and $t/J'=2.0$, where
the value of $J'/J$ may be appropriate for SrCu$_2$(BO$_3$)$_2$. 
The four-site unit cell of the Shastry-Sutherland lattice has 
eight different complex mean-field order parameters, as depicted in Fig.1.
Also, we assume the magnitude of the boson field $|b_i|$ to be spatially 
uniform. Notice that spatially inhomogeneous solution may appear at very 
low doping, but we do not consider this possibility here for simplicity. 

{\bf Zero temperature phase diagram}:
\begin{figure}[h]
\includegraphics[width=8.5cm]{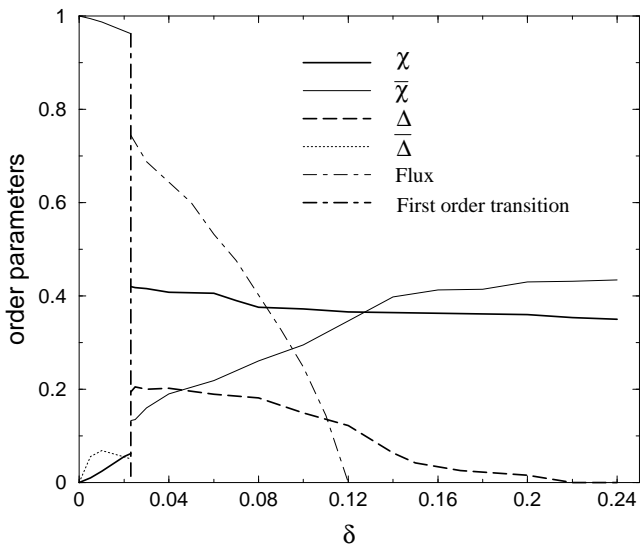}
\caption{Zero temperature phase diagram. The magnitudes of 
the order parameters $\chi$, $\bar{\chi}$, 
$\Delta$, $\bar{\Delta}$ (defined in the text and in unit of $t'$) 
are plotted as a function of 
doping $\delta$. 
The thin dot-dashed curve denoted by ``Flux'' is the value of
$\phi$ (phase factor) of the the plaquette product of $\chi$-fields  
(in unit of $\pi$). 
}
\label{T0phase}
\end{figure}
At $T=0$, all the bosons are condensed and there exist four different 
phases as shown in Fig.2. At the half-filling, the ground state is an 
insulator with the dimerization only on the diagonal bond:  
$\bar{\Delta}_1=\bar{\Delta}_2 \not= 0,$ others $= 0$  or 
$\bar{\chi}_1=\bar{\chi}_2 \not= 0,$ others $= 0$. 
Notice that the equivalence between the $\bar{\Delta}$ and 
$\bar{\chi}$ dimer states is a consequence of the SU(2) symmetry at 
the half-filling. This dimer ground state at the half-filling was 
also found in previous studies\cite{chung01,shastry02}. 
Away from half-filling, however, this symmetry is broken and
we find the phase with $|\bar{\chi}_1| = |\bar{\chi}_2| = \bar{\chi} 
\gg |\bar{\Delta}_{1}| = |\bar{\Delta}_2| = \bar{\Delta}$, 
$\bar{\chi} \gg |\chi_1|=|\chi_2| = \chi$, 
and $\Delta_1 = \Delta_2 = 0$ is the ground state up to $\delta < 0.023$. 
There is no flux from the $\chi$ field.
In this phase, since both $\bar{\Delta} \not= 0$ and 
$<b_{i}> \not= 0$, it is a superconducting phase 
where Cooper pairs are dimer singlets living on the 
diagonals\cite{shastry02}. 

At $\delta = 0.023$, there is a first order transition from the dimer phase 
to the staggered flux phase coexisting with $d$-wave superconductivity. 
In this phase, we have
$|\chi_1| = |\chi_2| = \chi$, 
$|\bar{\chi}_1| = |\bar{\chi}_2| = \bar{\chi}$,
$\Delta_1 = -\Delta_2 = \Delta$, and 
$\bar{\Delta}_1 = \bar{\Delta}_2 = 0$.
Moreover, the product of the $\chi$ fields has a non-trivial 
phase factor (or ``flux'') $\pm \phi$ on the plaquette and 
$\pm\phi /2$ on the triangle in a staggered fashion 
where $0 < \phi < \pi$. There exists a finite staggered orbital 
current circulating around the plaquette, and no current along the 
diagonals. It is worthwhile to notice that the staggered flux
phase coexisting with $d$-wave superconductivity was not 
discovered in the slave-boson mean-field theories of the
pure $t$-$J$ model on the square and triangular 
lattices\cite{palee92,kotliar88,subir00}
(except for the large-$N$ SU($N$) study where only the 
$\chi_{ij}$ field is considered\cite{marston88}).

As the doping is increased, the staggered flux order decreases
and vanishes at $\delta = 0.12$ where the ground state 
undergoes a continuous transition to the uniform $d$-wave 
superconducting state where $|\chi_1| = |\chi_2| = \chi$,
$|\bar{\chi}_1| = |\bar{\chi}_2| = \bar{\chi}$ with $\chi$ and 
$\bar{\chi}$ being comparable to each other, and   
$\Delta_1 = -\Delta_2 = \Delta$, 
$\bar{\Delta}_1 = \bar{\Delta}_2 = 0$.
There is no flux or orbital current in this phase.
This phase is similar to the $d$-wave superconducting phase found in 
the pure $t$-$J$ model on the square lattice\cite{palee92,kotliar88,subir00} 
except that we have nonzero diagonal $\bar{\chi}$ fields. 

When the doping reaches $\delta = 0.22$, the superconducting order
vanishes and the ground state becomes a normal Fermi liquid\cite{palee92,kotliar88}. 
In this phase, all the $\Delta$ fields are zero, and the $\chi$ fields are 
real with $\chi_1 = \chi_2 = \chi$, $\bar{\chi_1} = \bar{\chi_2} = \bar{\chi}$. 
Also, the magnitude of $\chi_1$ is comparable to that of $\bar{\chi_1}$. 

{\bf Finite temperature phase diagram}:
The phase diagram at finite temperatures as a function of 
doping and temperature (in unit of $t'$) is plotted in Fig.3.
The thick solid line represents the first order phase transition. 
To the right, finite $d$-wave order parameter $\Delta$ starts 
to develop below the dashed line ($T_{RVB1}$). 
Also, the $\chi_{ij}$ fields have a nontrivial flux below the 
long dashed line. To the left of the first order transition line, 
the particle-particle dimer order parameter $\bar{\Delta}$ is finite 
below the dot-dashed line ($T_{RVB2}$). Below the solid line ($T_{BEC}$), 
the bosons are condensed. The superconducting transition temperature 
$T_{sc}$ is the smaller of the $T_{RVB1}$ (or $T_{RVB2}$) and 
$T_{BEC}$\cite{palee92,kotliar88}. In order to get nonzero bose-condensate at finite 
temperatures, we introduced a small kinetic energy term along the
$c$-axis whose precise value is not important for the phase diagram.

\begin{figure}[h]
\includegraphics[width=8.5cm]{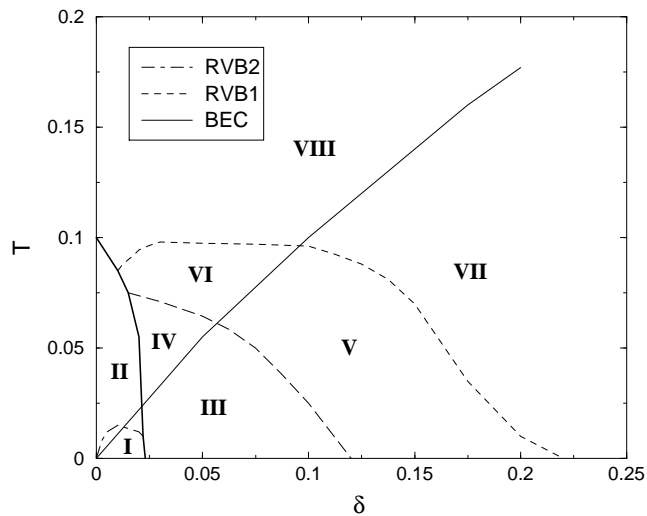}
\caption{Finite temperature phase diagram. The thick solid line 
represents the first order phase transition. The $d$-wave order 
parameter $\Delta$ is finite below the dashed 
line ($T_{RVB1}$). A nontrivial flux in $\chi_{ij}$ exists below the long 
dashed line. The dimer order parameter $\bar{\Delta}$ is finite 
below the dot-dashed line ($T_{RVB2}$). The bosons are condensed for $T<T_{BEC}$.
The superconducting transition temperature $T_{sc}$ is 
${\rm min} [T_{RVB1}$(or $T_{RVB2})$$,T_{BEC}]$. 
Here, the temperature is in unit of $t'$.
}
\label{Tphase}
\end{figure}

To the left of the first order transition line (thick solid line), 
the phase I represents the superconducting phase defined by 
$T < {\rm min}(T_{BEC},T_{RVB2})$ with ${\bar \Delta}\not=0$,
${\bar \chi}\not=0$, and $<b_i>\not=0$ and it is the same dimer 
superconductor discussed in the zero temperature case. 
The phase II can be a dimer phase without superconducting order 
when $T_{BEC} < T < T_{RVB2}$ with ${\bar \Delta}\not=0$, ${\bar \chi}\not=0$, 
and $<b_i>=0$. In the rest of the region II, there 
is no dimerization nor superconducting order.    
 
The phase III is the $d$-wave superconductor coexisting with the 
staggered flux order. Absence of the bose-condensation ($<b_i>=0$) in the 
phase IV leads to the staggered flux order without superconductivity, but with 
$d$-wave spin gap. As the doping is increased at low temperatures, 
the staggered flux order disappears and only pure $d$-wave superconductivity 
(with uniform $\chi$) remains in the phase V. Upon increasing temperature, 
superconductivity disappears in the phase VI, but $d$-wave spin gap survives.
The phase VII is the Fermi liquid phase and the phase VIII is the ``strange
metal'' phase with uniform $\chi$-field discussed in previous studies of
the $t$-$J$ model on the square and triangular lattices\cite{dhlee03,palee92}.  

{\bf Summary and Discussion}: 
We investigated possible superconducting phases and other competing orders
of the doped Mott insulator on the Shastry-Sutherland lattice within
the slave-boson mean-field theory of the $t$-$J$ model. One of the most
interesting discoveries is the $d$-wave superconducting 
state with coexisting staggered flux order or the staggered orbital current
at relatively lower doping as well as the pure $d$-wave superconductor
at higher doping. The coexisting phase was not discovered in the pure
$t$-$J$ model on the square and triangular lattices\cite{dhlee03,palee92}. 
Perhaps more interestingly, we found two different ``pseudogap'' phases;
one of them has the staggered flux order (staggered orbital current) and
the $d$-wave spin gap. The other has only the $d$-wave spin gap without
orbital current. It is intriguing to see the emergence of two different
pseudogap phases on the Shastry-Sutherland lattice in view of the fact
the long-range orbital antiferromagnetic order\cite{sudip01} 
and the $d$-wave spin gap with only the short-range
orbital antiferromagnetic correlations\cite{palee03} have been suggested as the 
origin of the pseudogap in cuprates. Clearly the geometric frustration is 
responsible for the complexity and competition shown in the phase diagram.
Whether the dynamic frustration due to the motion of holes in cuprates
leads to similar physics is an important question that needs to
be investigated in future. Moreover, our results provide directly 
relevant predictions for the doped version of the Mott-insulator 
SrCu$_2$(BO$_3$)$_2$ that may be available in future. 

This work was supported by the NSERC of Canada, Sloan Fellowship, 
and the CIAR. CHC and YBK thank the hospitality of the Aspen Center for 
Physics where some parts of this work were done 
during the Aspen summer workshop in 2003.

\end{document}